\documentclass[10pt,conference]{IEEEtran}

\usepackage{amsmath}
\usepackage{cite}
\usepackage{graphicx}
\usepackage{amsfonts}
\usepackage{latexsym}
\usepackage{subfigure}
\usepackage{algorithm}
\usepackage{algorithmic}
\usepackage{color}
\usepackage{times}
\usepackage{epsfig, graphics}
\usepackage{bm}
\usepackage{wrapfig,lipsum}

\begin{document}
\IEEEoverridecommandlockouts
\newtheorem{theorem}{Theorem}
\newtheorem{lemma}{Lemma}
\newtheorem{conjecture}{Conjecture}
\newtheorem{corollary}{Corollary}
\newtheorem{definition}{Definition}
\newtheorem{scheme}{Scheme}
\newcommand{\argmax}{\arg\!\max}
\newcommand{\rev}[1]{{\color{red}#1}} 
\newcommand{\pound}{\operatornamewithlimits{\gtrless}}

\title{Deep Learning for RF Signal Classification in Unknown and Dynamic Spectrum Environments 
\thanks{DISTRIBUTION A. Approved for public release: distribution unlimited.}
\thanks{\textsuperscript{\textcopyright} 2019 IEEE. Personal use of this material is permitted. Permission from IEEE must be
obtained for all other uses, in any current or future media, including 
reprinting/republishing this material for advertising or promotional purposes, creating new collective works, for resale or redistribution to servers or lists, or reuse of any copyrighted component of this work in other works.
}}

	\author{\IEEEauthorblockN{Yi Shi\IEEEauthorrefmark{1}, Kemal Davaslioglu\IEEEauthorrefmark{1}, Yalin E. Sagduyu\IEEEauthorrefmark{1}, 
			William C. Headley\IEEEauthorrefmark{2}, Michael Fowler\IEEEauthorrefmark{2}, and 
			Gilbert Green\IEEEauthorrefmark{3}
		}
		\IEEEauthorblockA{
			\IEEEauthorrefmark{1}Intelligent Automation, Inc., Rockville, MD, USA \\
			\IEEEauthorrefmark{2}Virginia Tech, Blacksburg, VA, USA \\
			\IEEEauthorrefmark{3}US Army, APG, MD, USA \\ 
	Email: \{yshi, kdavaslioglu, ysagduyu\}@i-a-i.com, \{cheadley, mifowler\}@vt.edu, gilbert.s.greeen2.civ@mail.mil		
	}
\vspace{-0.28in}
}

\maketitle

\begin{abstract}
Dynamic spectrum access (DSA) benefits from detection and classification of interference sources including in-network users, out-network users, and jammers that may all coexist in a wireless network. We present a deep learning based signal (modulation) classification solution in a realistic wireless network setting, where 1) signal types may change over time; 2) some signal types may be unknown for which there is no training data; 3) signals may be spoofed such as the smart jammers replaying other signal types; and 4) different signal types may be superimposed due to the interference from concurrent transmissions. For case 1, we apply continual learning and train a Convolutional Neural Network (CNN) using an Elastic Weight Consolidation (EWC) based loss. For case 2, we detect unknown signals via outlier detection applied to the outputs of convolutional layers using Minimum Covariance Determinant (MCD) and k-means clustering methods. For case 3, we extend the CNN structure to capture phase shifts due to radio hardware effects to identify the spoofing signal sources. For case 4, we apply blind source separation using Independent Component Analysis (ICA) to separate interfering signals. 
We utilize the signal classification results in a distributed scheduling protocol, where in-network (secondary) users employ signal classification scores to make channel access decisions and share the spectrum with each other while avoiding interference with out-network (primary) users and jammers. Compared with benchmark TDMA-based schemes, we show that distributed scheduling  constructed upon signal classification results provides major improvements to in-network user throughput and out-network user success ratio.
\end{abstract}
\vspace{-0.02225in}
\section{Introduction}

Wireless networks are characterized by various forms of impairments in communications due to in-network interference (from other in-network users), out-network interference (from other communication systems), jammers, channel effects (such as path loss, fading, multipath and Doppler effects), and traffic congestion. 
To support dynamic spectrum access (DSA), in-network users need to sense the spectrum and characterize interference sources hidden in spectrum dynamics.

\emph{Machine learning} provides automated means to classify received signals. Supported by recent computational and algorithmic advances, \emph{deep learning} is promising to extract and operate on latent representations of spectrum data that conventional machine learning algorithms have failed to achieve. In particular, deep learning can effectively classify signals based on their modulation types \cite{OShea2016,Kim16,Mendis16,Peng17,Ali17,Tu18,Shea18}.
Manifested in available datasets (e.g., \cite{OShea2016,Shea18}) for training wireless signal classifiers, a common practice in previous studies is to assume that signal types are known, remain unchanged, and appear without any interference and spoofing effects. However, those assumptions are typically invalid in a realistic wireless network, where 
\begin{enumerate}
\item signal types change over time;
\item some signal types are not known \emph{a priori} and therefore there is no training data available for those signals; 
\item signals are potentially spoofed, e.g., a smart jammer may replay received signals from other users thereby hiding its identity; and 
\item  signals are superimposed due to the interference effects from concurrent transmissions of different signal types.
\end{enumerate}
\begin{figure*}
	\centering
	\includegraphics[width=1.7\columnwidth]{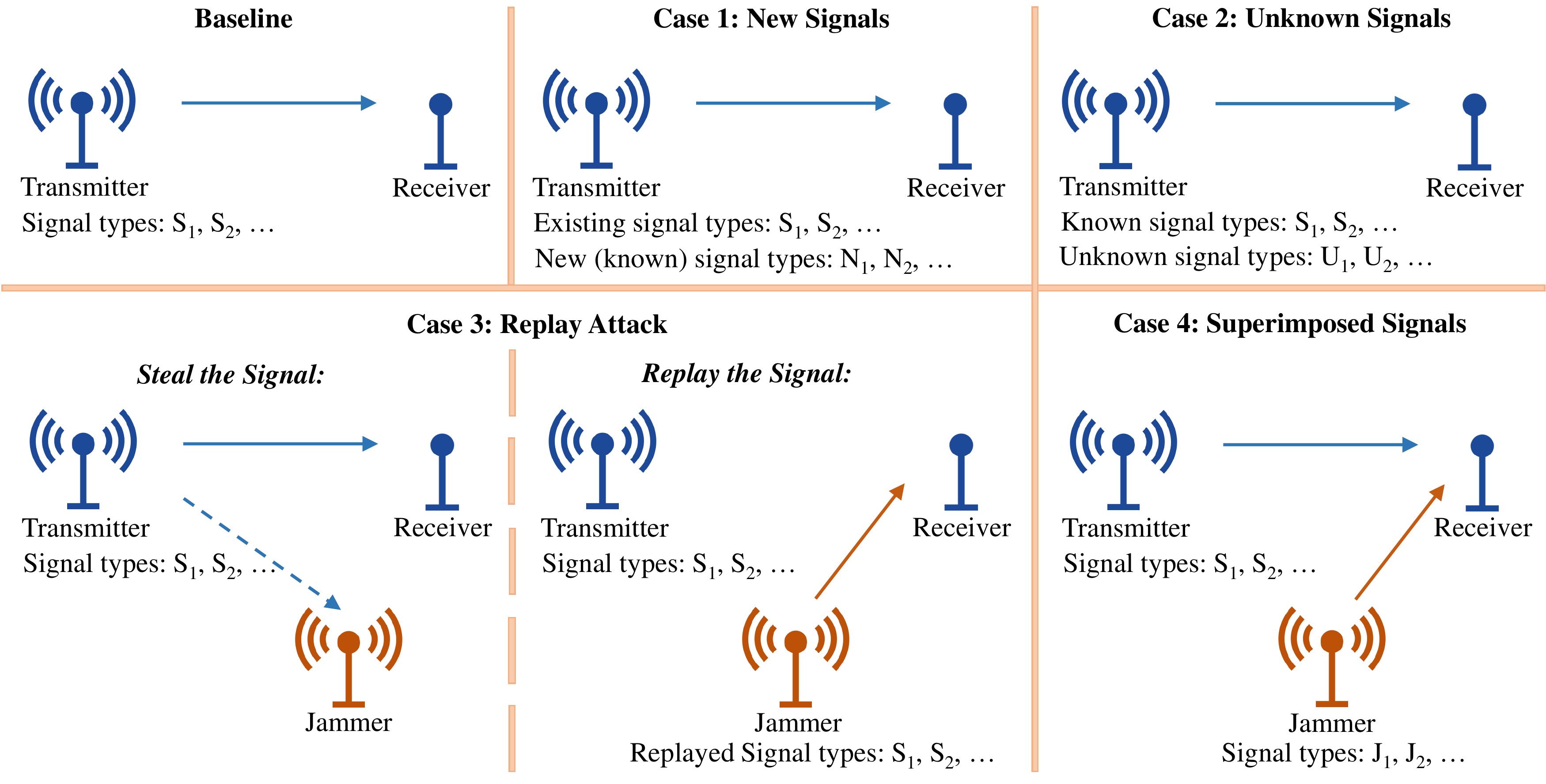}
	\caption{RF signal classification cases, including new signals, unknown signals, replay attacks from jammers, and superimposed signals. }
	\label{fig:hor_cases1-4}
\end{figure*}
It is essential to incorporate these four realistic cases (illustrated in Fig. \ref{fig:hor_cases1-4}) in building the RF signal classifier so that its outcomes can be practically used in a DSA protocol. We consider a wireless signal classifier that classifies signals based on modulation types into \emph{idle}, \emph{in-network users} (such as secondary users), \emph{out-network users} (such as primary users), and \emph{jammers}. Using the signal classification results, in-network users allocate  time slots for collision-free scheduling in a distributed setting and share the spectrum with each other while protecting out-network user transmissions and avoiding interference from jammers. 



Assuming that different signal types use different modulations, we present a convolutional neural network (CNN) that classifies the received I/Q samples as idle, in-network signal, jammer signal, or out-network signal. We start with the simple baseline scenario that all signal types (i.e., modulations) are fixed and known (such that training data are available) and there are no superimposed signals (i.e., signals are already separated). The average accuracy over all signal-to-noise-ratios (SNRs) is $0.934$.
We then extend the signal classifier to operate in a realistic wireless network as follows.
\begin{enumerate}
\item New modulations appear in the network over time (see case 1 in Fig. \ref{fig:hor_cases1-4}) and should be classified as specified signal types. Instead of retraining the signal classifier, we design a \emph{continual learning} algorithm \cite{Ring94} to update the classifier with much lower cost, namely by using  an  Elastic  Weight  Consolidation  (EWC). This approach achieves over time the level of performance similar to the ideal case when there are no new modulations.

\item Some signal types such as modulations used in jammer signals are unknown (see case 2 in Fig. \ref{fig:hor_cases1-4}) such that there is no available training data for supervised learning. We present an \emph{outlier detection} solution to achieve high accuracy in classifying signals with unknown jamming signals. For that purpose, we apply Minimum  Covariance  Determinant  (MCD) and  k-means clustering  methods at the outputs of the signal classifier's  convolutional layers. This approach successfully classifies all inliers and most of outliers, achieving $0.88$ average accuracy.

\item  Smart jammers launch \emph{replay attacks} by recording signals from other users and transmitting them as jamming signals (see case 3 in Fig. \ref{fig:hor_cases1-4}). We extend  the  CNN  structure to  capture  phase  shift  due  to  radio  hardware  effects  to  identify the  spoofing  signals  and  relabel  them  as  jammers. This approach achieves $0.972$ accuracy in classifying superimposed signals.

\item Wireless signals are received as \emph{superimposed} (see case 4 in Fig. \ref{fig:hor_cases1-4}) if transmitted at the same time (on the same frequency). We apply blind source  separation using Independent  Component  Analysis  (ICA) \cite{Amari96} to obtain each single signal that is further classified by deep learning. This approach achieves $0.837$ average accuracy.
\end{enumerate}

The signal classification results are used in the DSA protocol that we design as a \emph{distributed scheduling} protocol, where an in-network user transmits if the received signal is classified as idle or in-network (possibly superimposed). If the received signal is classified as in-network, the in-network user needs to share the spectrum with other in-network user(s) based on the confidence of its classification. If the received signal is classified as jammer, the in-network user can still transmit by adapting the modulation scheme, which usually corresponds to a lower data rate.
We assume that a transmission is successful if the signal-to-interference-and-noise-ratio (SINR) at the receiver is greater than or equal to some threshold required by a modulation scheme.
If out-network signals are detected, the in-network user should not transmit to avoid any interference, i.e., out-network users are treated as primary users. Results show that this approach achieves higher throughput for in-network users and higher success ratio for our-network users compared with benchmark (centralized) TDMA schemes.

The rest of the paper is organized as follows.
Section~\ref{sec:relatedwork} discusses related work.
Section~\ref{sec:sam} presents the deep learning based signal classification in unknown and dynamic spectrum environments.
Section~\ref{sec:scheduling} introduces the distributed scheduling protocol as an application of deep learning based spectrum analysis.
Section~\ref{sec:conclusion} concludes the paper.


\section{Related Work} \label{sec:relatedwork}

Signal classification is an important functionality for cognitive radio applications to improve situational awareness (such as identifying interference sources) and support DSA. The traditional approaches for signal classification include likelihood based methods or feature based analysis on the received I/Q samples \cite{Dobre07,Dobre05,Headley11}. However, these two approaches require expert design or knowledge of the signal. They also add complexity to a receiver since the raw I/Q data must be manipulated before classification. 

By learning from spectrum data, machine learning has found rich applications in wireless communications \cite{Alsheikh2014, Chen2017}. In particular, deep learning has been applied to learn complex spectrum environments, including spectrum sensing by a CNN \cite{Lee2017}, spectrum data augmentation by generative adversarial network (GAN) \cite{Kemal2018, Terpek18}, channel estimation by a feedforward neural network (FNN) \cite{DeepOFDM}, and jamming/anti-jamming with FNN in training and test times \cite{Shi18,Yi18, Secon2019}. Modulation classification has been extensively studied  with deep neural networks \cite{OShea2016,Kim16,Mendis16,Peng17,Ali17,Tu18}, where the goal is to classify a given isolated signal to a known modulation type. In \cite{Shea18}, the performance of modulation classification was evaluated with over-the-air measurements. 
\cite{deVrieze18} shows that in performance real-world systems depends on whether the training dataset fully captures the variety of interference and hardware types in the real radio environment. Those approaches cannot be readily applied in a wireless network setting, as they do not capture dynamic and unknown signal types, smart jammers that may spoof signal types (e.g., signals may be generated through the GAN \cite{shi2019generative}) and superposition of signals types due to concurrent transmissions. In this paper, we address these issues to make signal classification applicable for use in a DSA protocol.

\section{Deep Learning based Spectrum Analysis}
\label{sec:sam}


We consider different modulation schemes used by different types of users transmitting on a single channel. We start with the baseline case where modulations used by different user types are known and there is no signal superposition (i.e., interfering sources are already separated). We categorize modulations into four signal types:
\begin{enumerate}
\item idle: no signal
\item in-network user signals: QPSK, 8PSK, CPFSK
\item jamming signals: QAM16, QAM64, PAM4, WBFM
\item out-network user signals: AM-SSB, AM-DSB, GFSK
\end{enumerate}
There are in-network users (trying to access the channel opportunistically), out-network users (with priority in channel access) and jammers that all coexist. Out-network users are treated as primary users and their communications should be protected. Without prior domain knowledge other than training data, an in-network user classifies received signals to idle, in-network, jammer, or out-network. 
The classifier computes a score vector $(p_0,p_{in}, p_{jam},$ $p_{out})$ for each instance, where $p_0$, $p_{in}$, $p_{jam}$, and $p_{out}$ are the likelihood scores for classifying signals as idle, in-network, jammer, and out-network, respectively. If one score is larger than the other three, the instance is classified as the corresponding case.


\subsection{The Classifier Structure and Performance}
\label{subsec:structure}

We use the dataset in \cite{OShea2016}. Each sample in the dataset consists of $128$ complex valued data points, i.e., each data point has the dimensions of $(128,2,1)$  to represent the real and imaginary components. We use $10$ modulations (QPSK, 8PSK, QAM16, QAM64, CPFSK, GFSK, PAM4, WBFM, AM-SSB, and AM-DSB) collected over a wide range of SNRs from -20~dB to 18~dB in 2~dB increments. These modulations are categorized into signal types as discussed before. At each SNR, there are 1000~samples from each modulation type. Instead of using a conventional feature extraction or off-the-shelf deep neural network architectures such as ResNet, we build a custom deep neural network that takes I/Q data as input. 

\begin{figure}
	\centering
	\includegraphics[width=1.0\columnwidth]{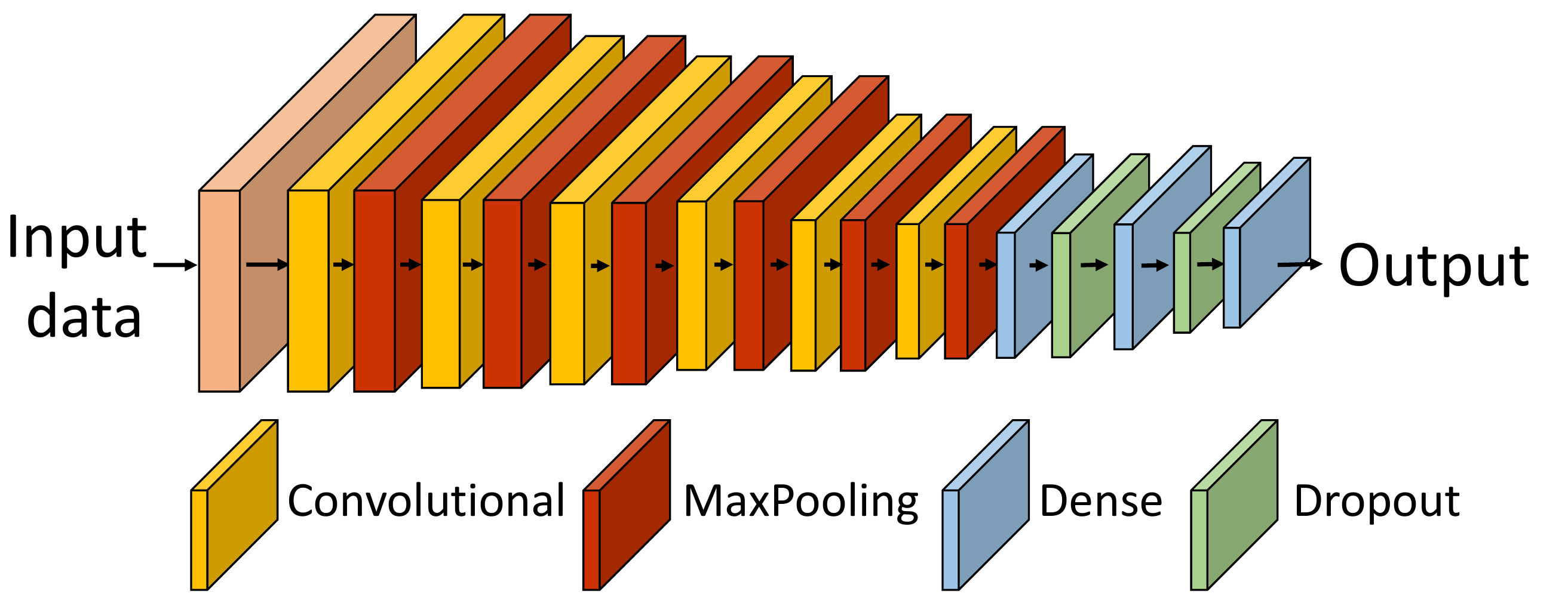}
	\caption{CNN classifier structure for RF signal classification.}
	\label{fig:cnn}
\end{figure}

We train a CNN classifier that consists of several convolutional layers and fully connected layers in the last three stages. 
However, when the filter size in the convolutional layers is not divisible by the strides, it can create checkerboard effects (see \cite{Odena16} for more details). In the CNN classifier structure, shown in Fig.~\ref{fig:cnn}, we paid attention to avoid the checkerboard effects and used the following layers:
\begin{itemize}
\item Input shape: $(128,2)$
\item 2D ZeroPadding with size $(1,1)$
\item Convolutional layer with $128$ filters with size of $(3,3)$
\item 2D MaxPolling layer with size $(2,1)$ and stride $(2,1)$
\item Five cascades of the following:
    \begin{itemize}
	\item 2D Zeropadding with size $(1,1)$
	\item Convolutional layer with $256$ filters with size of $(3,3)$ 
	\item 2D MaxPolling layer with pool size $(2,2)$ and stride $(2,1)$
    \end{itemize}
\item Fully connected layer with $256$ neurons and Scaled Exponential Linear Unit (SELU) activation function, which is $x$ if $x>0$ and $ae^x-a$ if $x \le 0$ for some constant $a$
\item Dropout with probability $0.5$
\item Fully connected layer with $64$ neurons and SELU activation function
\item Dropout with probability $0.5$
\item Fully connected layer with $4$ neurons and SELU activation function
\end{itemize}

The classifier is trained in TensorFlow \cite{Tensorflow}. The ADAM optimizer \cite{Adam} is used with a step size of $5 \times 10^{-5}$ and the categorical cross-entropy loss function is used for training. Cross-entropy function is given by
\begin{align}
\mathcal{L}(\boldsymbol{\theta}) =  -\sum_{i} \beta_i \log(y_i),
\end{align}
where $\boldsymbol{\theta}$ is the set of the neural network parameters and $\{\beta_i\}_{i=1}^m$ is a binary indicator of ground truth such that $\beta_i=1$ only if $i$ is the correct label among $m$ classes (labels). The neural network output $\textbf{y} \in \mathrm{R}^m$ is an $m$-dimensional vector, where each element in $y_i \in \textbf{y}$ corresponds to the likelihood of that class being correct. 
\begin{figure}[t]
	\centering
	\includegraphics[trim={0 0.75cm 0 1cm},clip, width=1.0\columnwidth]{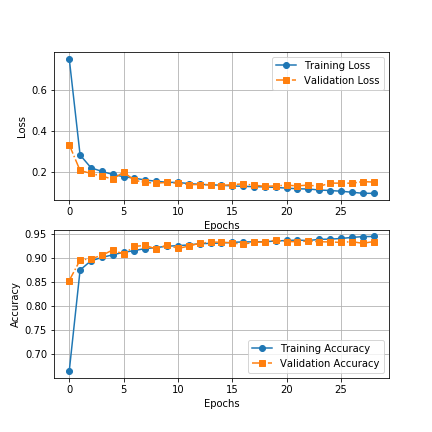}
	\caption{CNN classifier performance.}
	\label{fig:lossacc}
\end{figure}

\begin{figure}
	\centering
	\includegraphics[width=0.75\columnwidth]{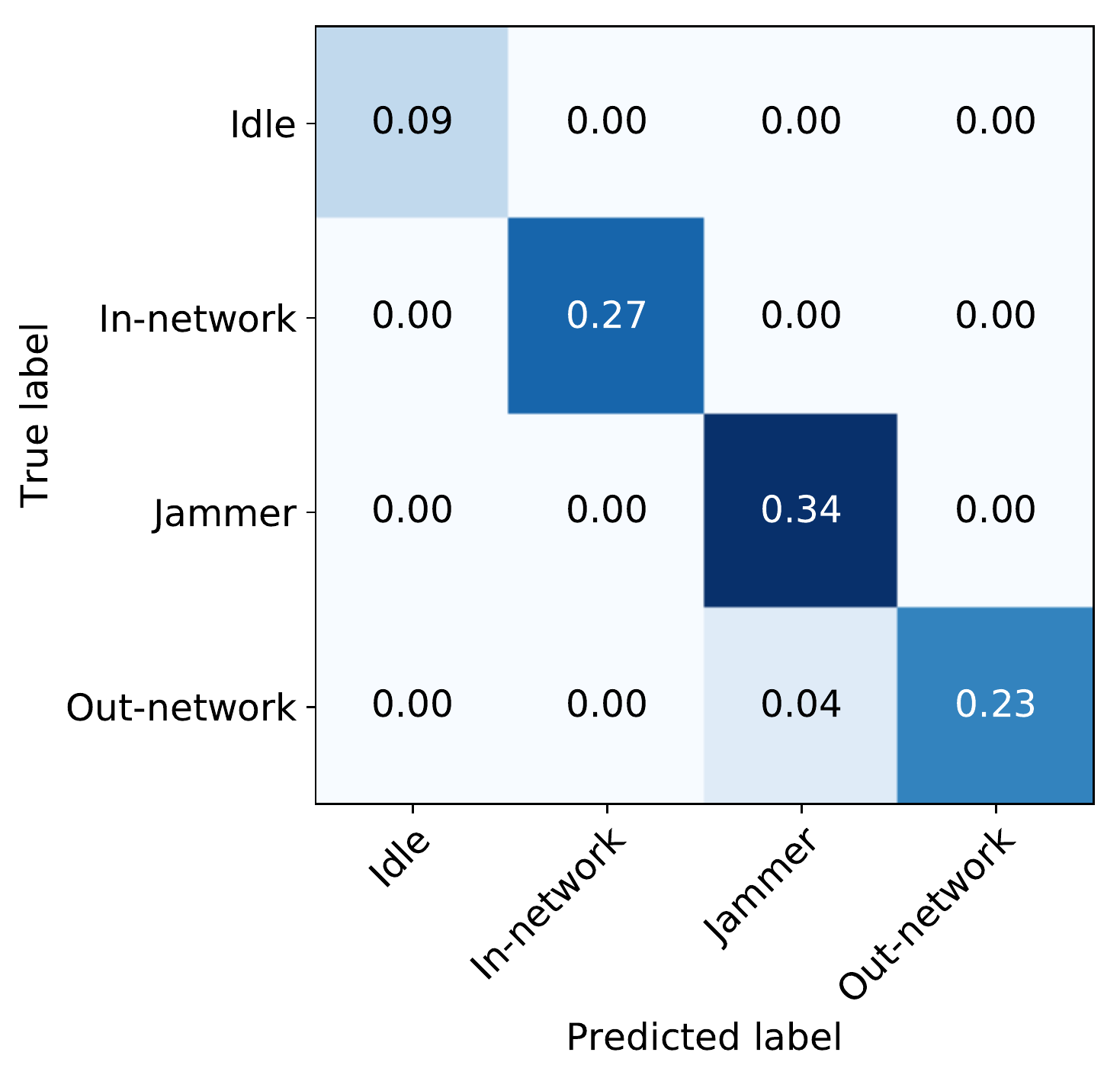}
	\caption{Confusion matrix (averaged over all SNRs).}
	\label{fig:confusion1}
\end{figure}

\begin{figure*}
	\centering
	\subfigure[0 dB.]
	{\includegraphics[width=0.6\columnwidth]{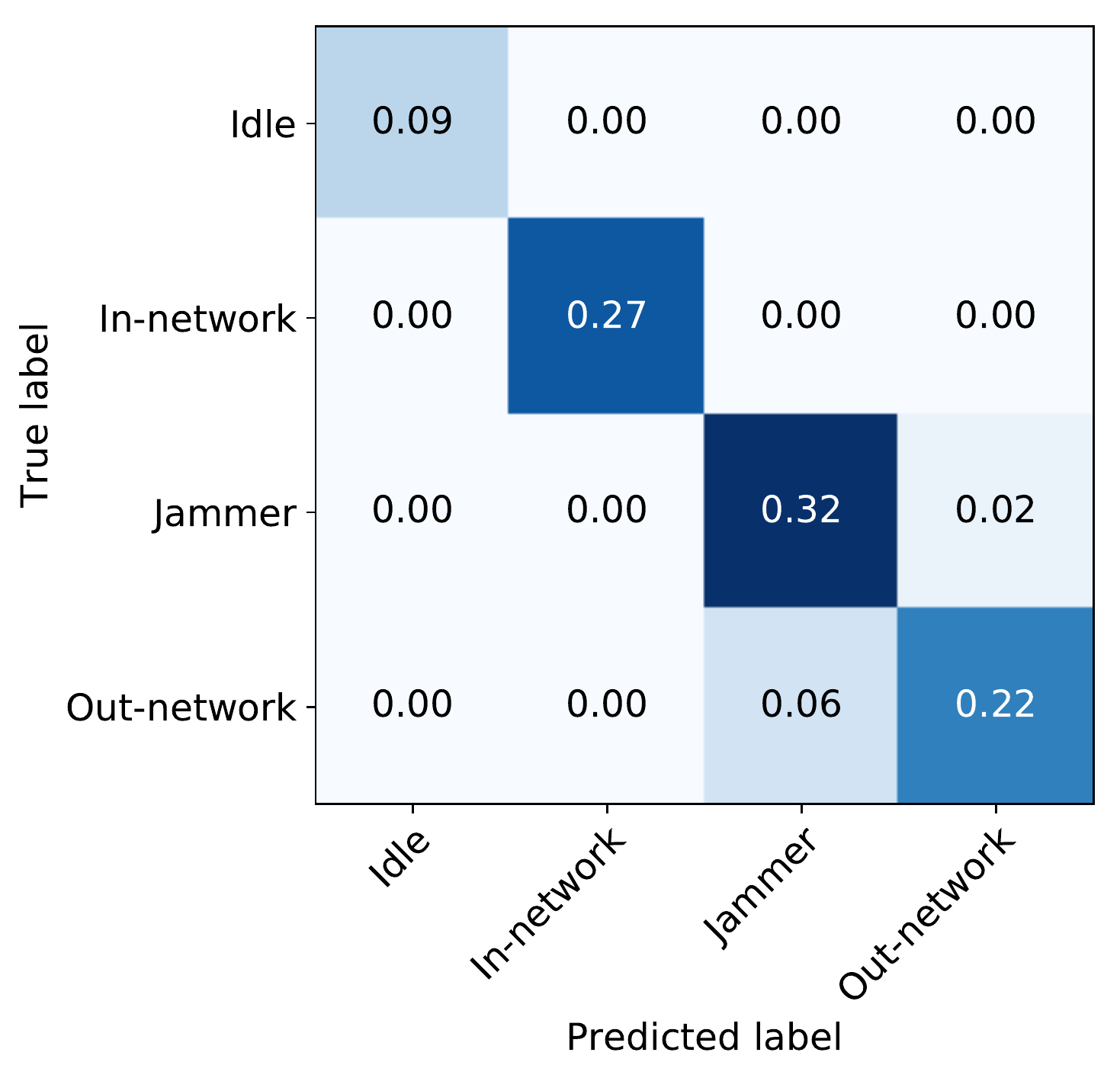}}
	\hfil
	\subfigure[10 dB.]
	{\includegraphics[width=0.6\columnwidth]{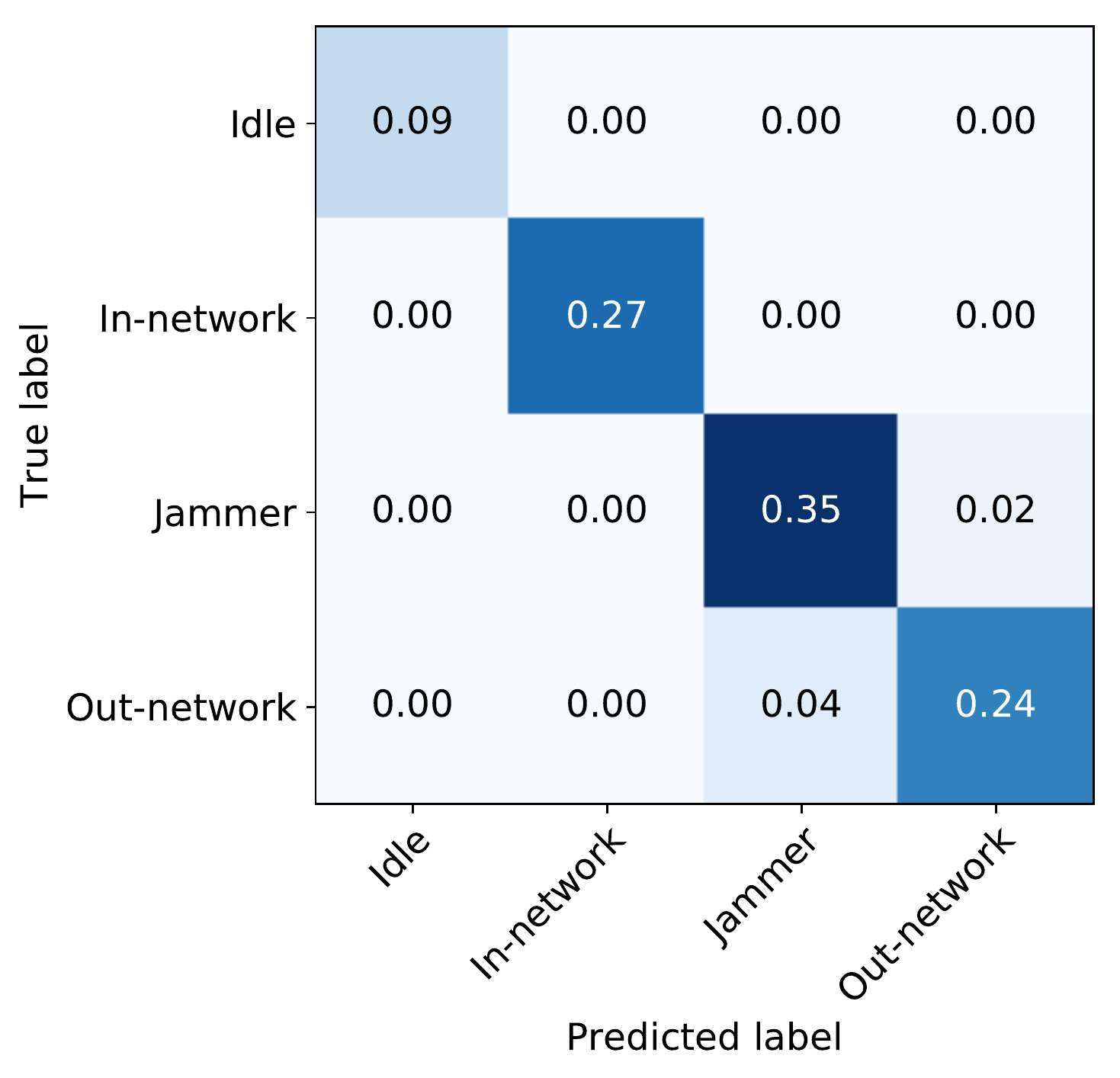}}
	\hfil
	\subfigure[18 dB.]
	{\includegraphics[width=0.6\columnwidth]{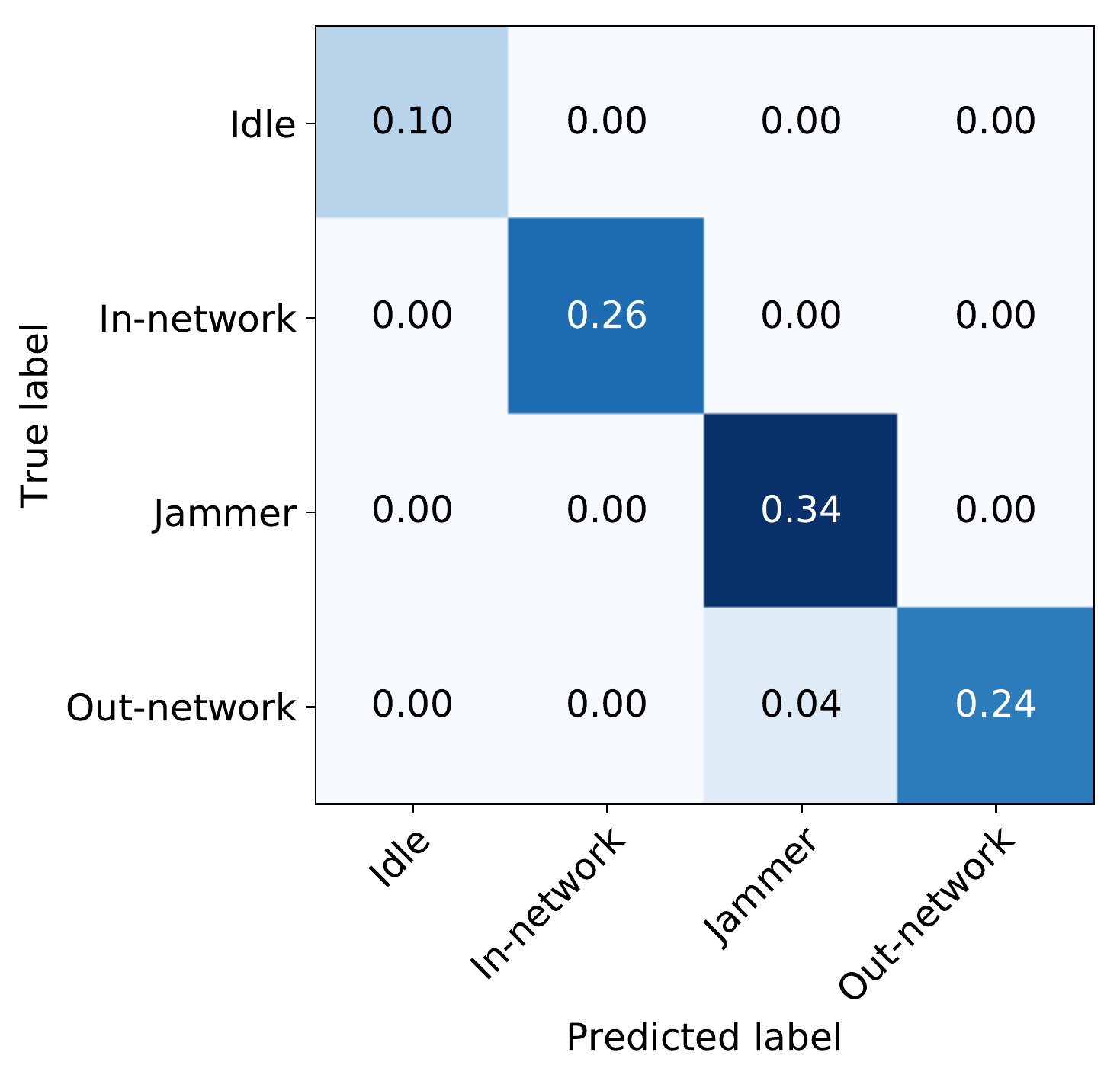}}
	\caption{Confusion matrices at different SNR values.}
\label{fig:confusion2}
\end{figure*}

We split the data into $80\%$ for training and $20\%$ for testing. The loss function and accuracy are shown in Fig.~\ref{fig:lossacc} as a function of training epochs. The testing accuracy is $0.934$. Fig.~\ref{fig:confusion1} shows the average confusion matrix of the classifier over all SNR levels. 
Table~\ref{fig:accuracy} shows the average accuracy vs. SNR over all types of signals. The SNR levels are from $0$ to $18$dB in $2$dB increments. Fig.~\ref{fig:confusion2} shows confusion matrices at $0$dB, $10$dB, and $18$dB SNR levels.



\begin{table}
	\caption{CNN classifier accuracy (averaged over all signal types).}
	\centering
	{\small
		\begin{tabular}{c|c||c|c}
			SNR (dB) & Accuracy & SNR (dB) & Accuracy \\ \hline
			0 & 0.906 & 10 & 0.942 \\ 
			2 & 0.930 & 12 & 0.950 \\
			4 & 0.928 & 14 & 0.951 \\
			6 & 0.933 & 16 & 0.933 \\
			8 & 0.934 & 18 & 0.934 \\ \hline
		\end{tabular}
	}
	\label{fig:accuracy}
\end{table}


\subsection{Continual Learning}
\begin{figure}
	\centering
	\includegraphics[width=1.0\columnwidth]{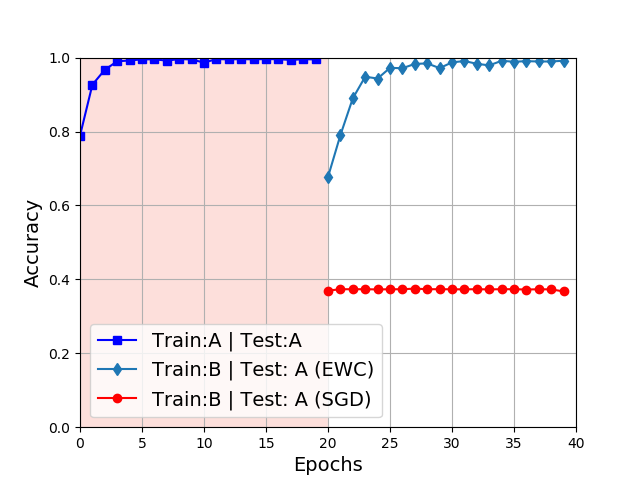}
	\caption{Classifier performance over time.}
	\label{fig:perftime}
\end{figure}

So far, we assumed that all modulation types are available in training data.
We now consider the case that initially five modulations are taught to the classifier. Over time, three new modulations are introduced. Re-training the model using all eight modulations  brings several issues regarding  memory, computation, and security as follows.
\begin{itemize}
\item \textit{Memory}: Previous data needs to be stored.
\item \textit{Computation}: Retraining using the complete dataset will take longer.
\item \textit{Security}: If a device or server is compromised, adversary will have the data to train its own classifier, since previous and new data are all stored.
\end{itemize}
On the other hand, if a model is re-trained using the new three modulations with Stochastic Gradient Descent (SGD), performance on the previous five modulations drops significantly (see Fig.~\ref{fig:perftime}). This is called \emph{catastrophic forgetting} \cite{GoodfellowCatastrophic,EWC17}. In Fig.~\ref{fig:perftime}, Task A is the classification of first five modulations and Task B is the classification of the next three new modulations. SGD suffers from catastrophic forgetting and its accuracy on Task A drops to $0.37$ when retrained with Task B.

We apply EWC to address this problem. EWC slows down learning on selected neural network weights to remember previously learned tasks (modulations) \cite{EWC17}. EWC augments loss function using Fisher Information Matrix that captures the similarity of new tasks and uses the augmented loss function $L(\theta)$ given by 
\begin{eqnarray}
L(\theta)=L_B (\theta)+ \sum_i \frac{\lambda}{2} F_i (\theta_i - \theta_{A,i}^* )^2 \; ,
\end{eqnarray}
where $\theta_A$ denotes the weights used to classify the first five modulations (Task A), $L_B (\theta)$ is the loss function for Task B, $F_i$ is the fisher information matrix that determines the importance of old and new tasks, and $i$ denotes the parameters of a neural network. Higher values on the Fisher diagonal elements $F_i$ indicate more certain knowledge, and thus they are less flexible. This approach helps identify and protect weights. In Fig.~\ref{fig:perftime}, we can see that EWC mitigates catastrophic learning to improve the accuracy on Task B such that the accuracy increases over time to the level of Task A.

\subsection{Classifier for Unknown Type of Signals}

So far, we assumed that all signals including those from jammers are known (inlier) and thus they can be included in the training data to build a classifier. This assumption is reasonable for in-network and out-network user signals. However, jamming signals are possibly of an unknown type (outlier). Then a classifier built on known signals cannot accurately detect a jamming signal. An \emph{outlier detection} is needed as a robust way of detecting if the (jamming) signal is known or unknown. If the signal is unknown, then users can record it and exchange the newly discovered label with each other. If the signal is known, then the signal passes through the classifier to be labeled. For the outlier detection, as the waveform dimensions are large, we reuse the convolutional layers of the classifier to extract the features of the received signal. Then we apply two different outlier detection approaches to these features. 

\subsubsection{MCD-based Classifier}

The first method for the outlier detection is based on the Minimum Covariance Determinant (MCD) method \cite{MCD,MCD3}. MCD fits an elliptic envelope to the test data such that any data point outside the ellipse is considered as an outlier. MCD uses the Mahalanobis distance to identify outliers:
\begin{eqnarray}
MD(\mathbf{x})= \sqrt{(\mathbf{x} - \mu_x)^T (S_x)^{-1} (\mathbf{x} - \mu_x)} \; ,
\end{eqnarray}
where $\mu_x$ and $S_x$ are the mean and covariance of data $x$, respectively. We tried two approaches: i) directly apply outlier detection using MCD and ii) extract features and apply MCD outlier detection to these features.

The evaluation settings are as the following:
\begin{itemize}
\item Inlier signals: QPSK, 8PSK, CPFSK, AM-SSB, AM-DSB, GFSK
\item Outlier signals: QAM16, QAM64, PAM4, WBFM
\end{itemize}
The second approach of feature extraction followed by outlier detection yields the best performance. In the feature extraction step, we freeze the model in the classifier and reuse the convolutional layers. The output of convolutional layers in the frozen model are then input to the MCD algorithm. MCD algorithm has a variable called contamination that needs to be tuned. Contamination accounts for the estimated proportion of outliers in the dataset. In the training step of MCD classifier, we only present the training set of known signals (in-network and out-network user signals), while in the validation step, we test the inlier detection accuracy with the test set of inliers and test the outlier detection accuracy with the outlier set (jamming signals). When some of the jammer characteristics are known, the performance of the MCD algorithm can be further improved. Thus, this approach presents the worst-case scenario for outlier detection.

\begin{figure}
	\centering
	\includegraphics[width=0.95\columnwidth]{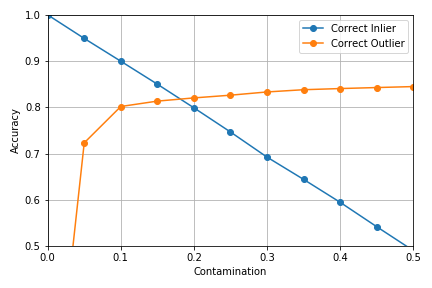}
	\caption{MCD-based classifier accuracy on inliers and outliers.}
	\label{fig:inoutaccuracy}
\end{figure}

The classification accuracy for inliers and outliers as a function of contamination factor in MCD is shown in Fig.~\ref{fig:inoutaccuracy}. The best contamination factor is $0.15$, which maximizes the minimum accuracy for inliers and outliers.

\begin{figure*}
	\centering
	\subfigure[0 dB.]
	{\includegraphics[width=0.6\columnwidth]{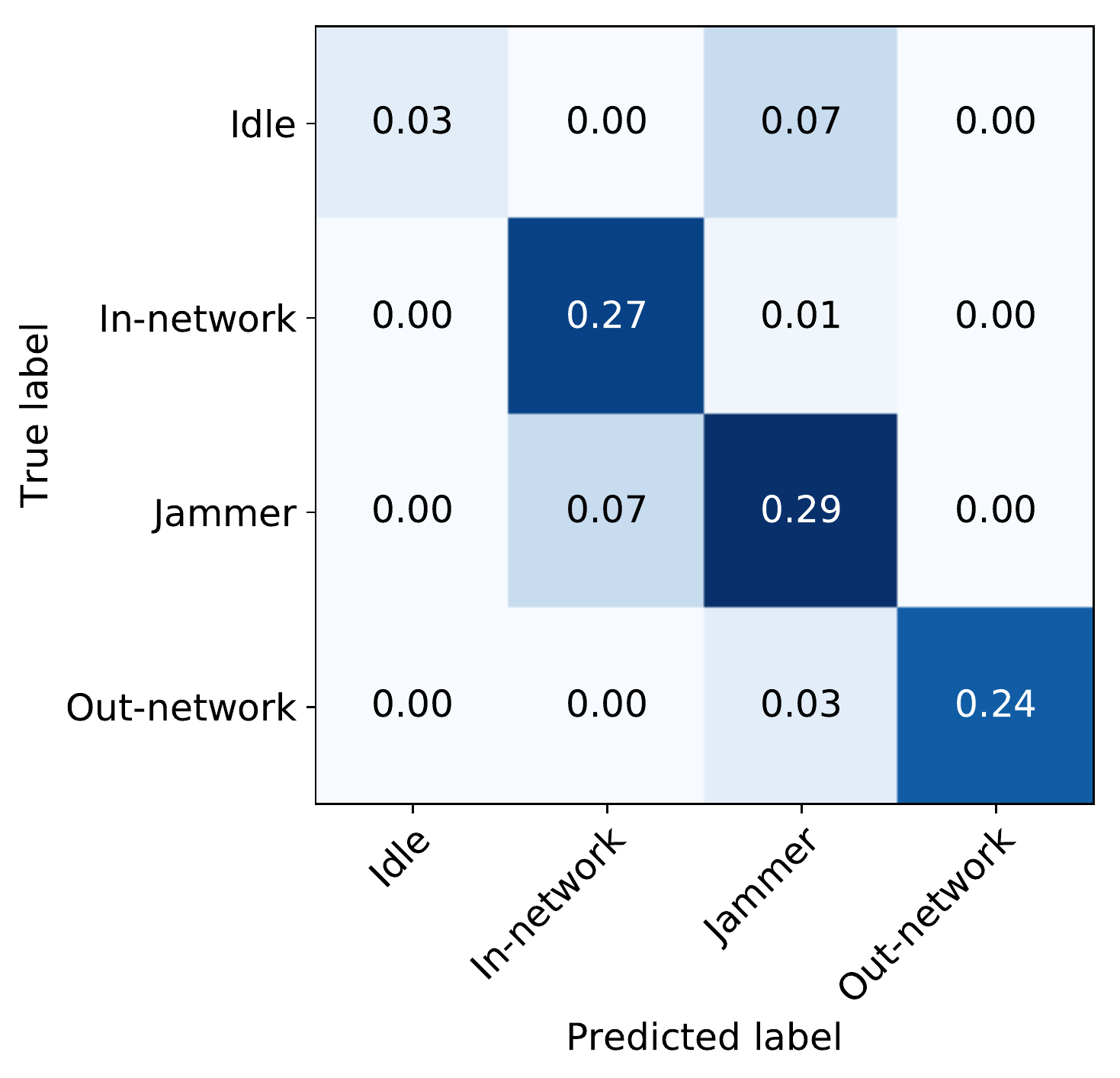}}
	\hfil
	\subfigure[10 dB.]
	{\includegraphics[width=0.6\columnwidth]{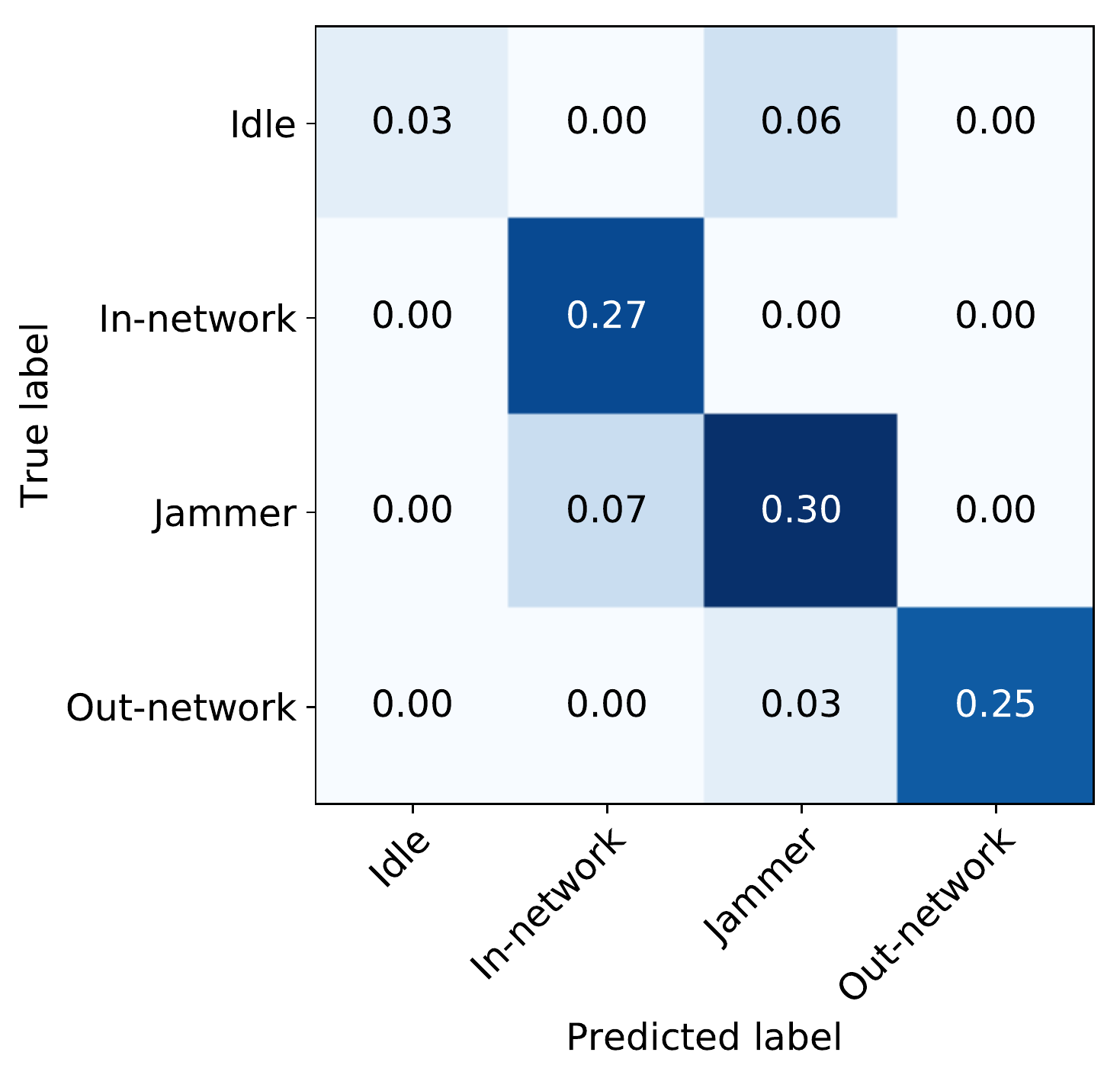}}
	\hfil
	\subfigure[18 dB.]
	{\includegraphics[width=0.6\columnwidth]{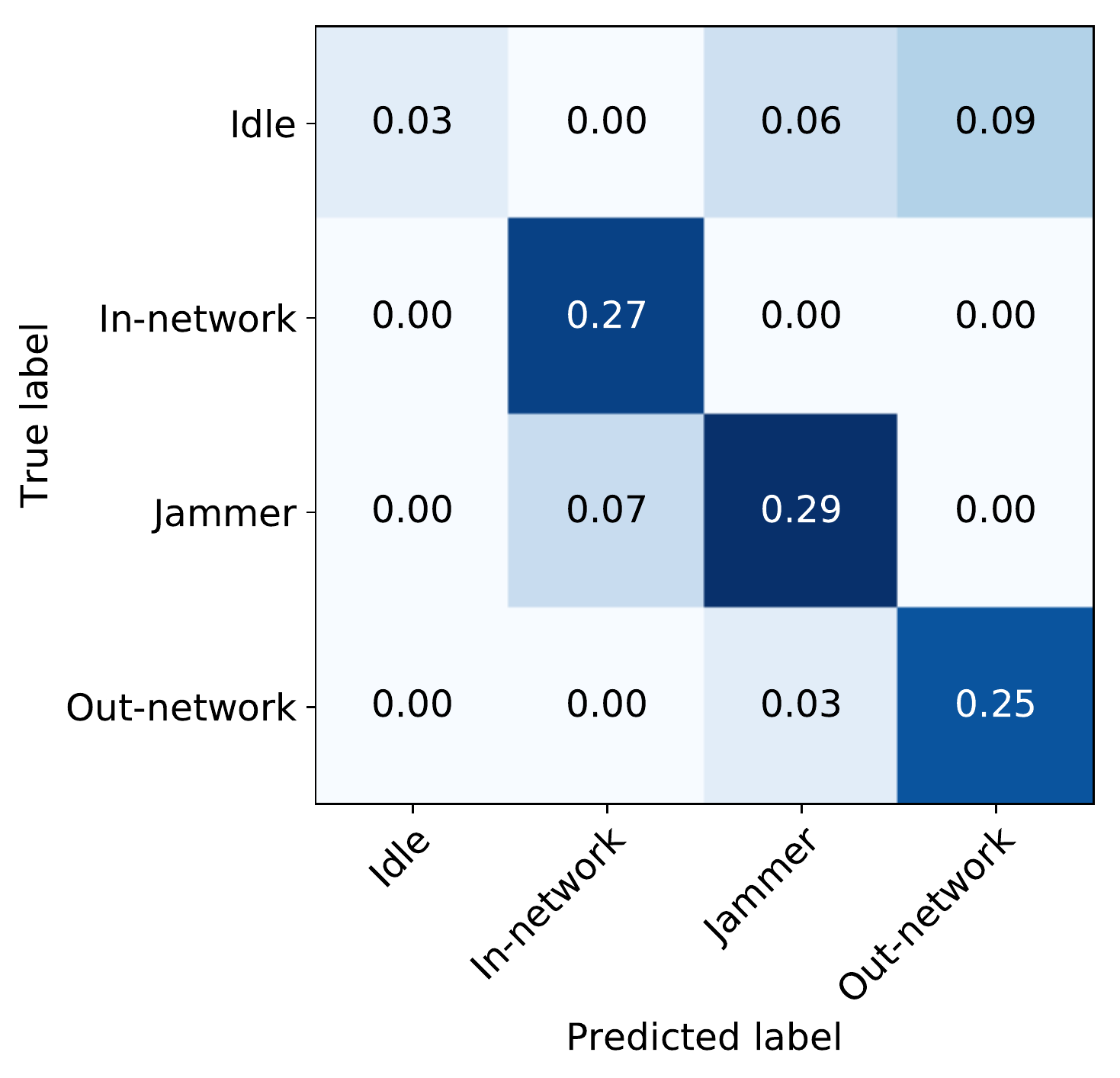}}
	\caption{MCD-based outlier detection confusion matrices at different SNR values.}
\label{fig:confusion3}
\end{figure*}

Table~\ref{table:mcdaccuracy} shows the accuracy as a function of SNR and Fig.~\ref{fig:confusion3} shows confusion matrices at $0$dB, $10$dB, and $18$dB SNR levels.


\begin{table}
	\caption{MCD-based outlier detection accuracy over all SNR values.}
	\centering
	{\small
		\begin{tabular}{c|c||c|c}
			SNR (dB) & Accuracy & SNR (dB) & Accuracy \\ \hline
			0 & 0.822 & 10 & 0.843 \\ 
			2 & 0.814 & 12 & 0.844 \\
			4 & 0.824 & 14 & 0.847 \\
			6 & 0.832 & 16 & 0.844 \\
			8 & 0.845 & 18 & 0.839 \\ \hline
		\end{tabular}
	}
	\label{table:mcdaccuracy}
\end{table}


\subsubsection{k-means Clustering based Classifier}

\begin{figure}
	\centering
	\includegraphics[width=0.4\columnwidth]{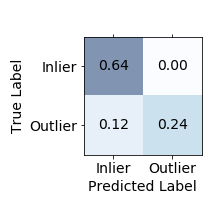}
	\caption{Confusion matrix for k-means clustering based outlier detection.}
	\label{fig:kmeans}
\end{figure}

The second method for the outlier detection is the k-means clustering method. This method divides the samples into $k=2$ clusters by iteratively finding $k$ cluster centers. We again have in-network and out-network user signals as inlier and jamming signals as outlier. We first use CNN to extract features and then use k-means clustering to divide samples into two clusters, one for inlier and the other for outlier. The confusion matrix is shown in Fig.~\ref{fig:kmeans}. The accuracy of correctly identifying inliers has improved with k-means compared to the MCD method. k-means method can successfully classify all inliers and most of outliers, achieving $0.88$ average accuracy.

\subsection{Detection of a Smart Jammer}
Next, we consider a smart jammer that records an in-network user signal, and then amplifies and forwards it as a replay attack (instead of transmitting a distinct jamming signal, as assumed before). Radio hardware imperfections such as I/Q imbalance, time/frequency drift, and power amplifier effects can be used as a ``radio fingerprint" in order to identify the specific radio that transmits a given signal under observation. In particular, we aim to design a classifier using I/Q data with hardware impairments to identify the type of a transmitter (in-network user or jammer).

Suppose the jammer receives the in-network user signal, which is QAM64 at $18$ dB SNR, and collects $1000$ samples. Then the jammer amplifies and forwards it for jamming. We model the hardware impairment as a rotation on the phase of original signal. This offset will be used in the classifier to detect a jamming signal in a replay attack.


The jammer rotates $1000$ samples with different angles $\theta = \frac{k \pi}{16}$ for $k=0,1, \cdots,16$. 
The jammer uses these signals for jamming. Each of these signals has its $e^{j \theta}$ rotation.


\begin{figure}
	\centering
	\subfigure[Loss value.]
	{\includegraphics[width=0.8\columnwidth]{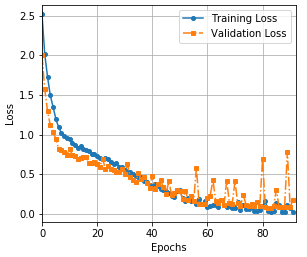}}
	\hfil
	\subfigure[Accuracy.]
	{\includegraphics[width=0.8\columnwidth]{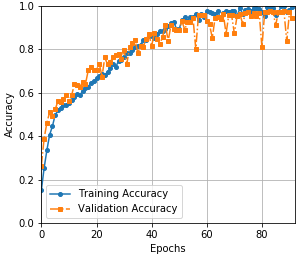}}
	\caption{Classifier performance to detect $17$ types of signals.}
\label{fig:17signals}
\end{figure}

\begin{figure}
	\centering
	\includegraphics[width=1.0\columnwidth]{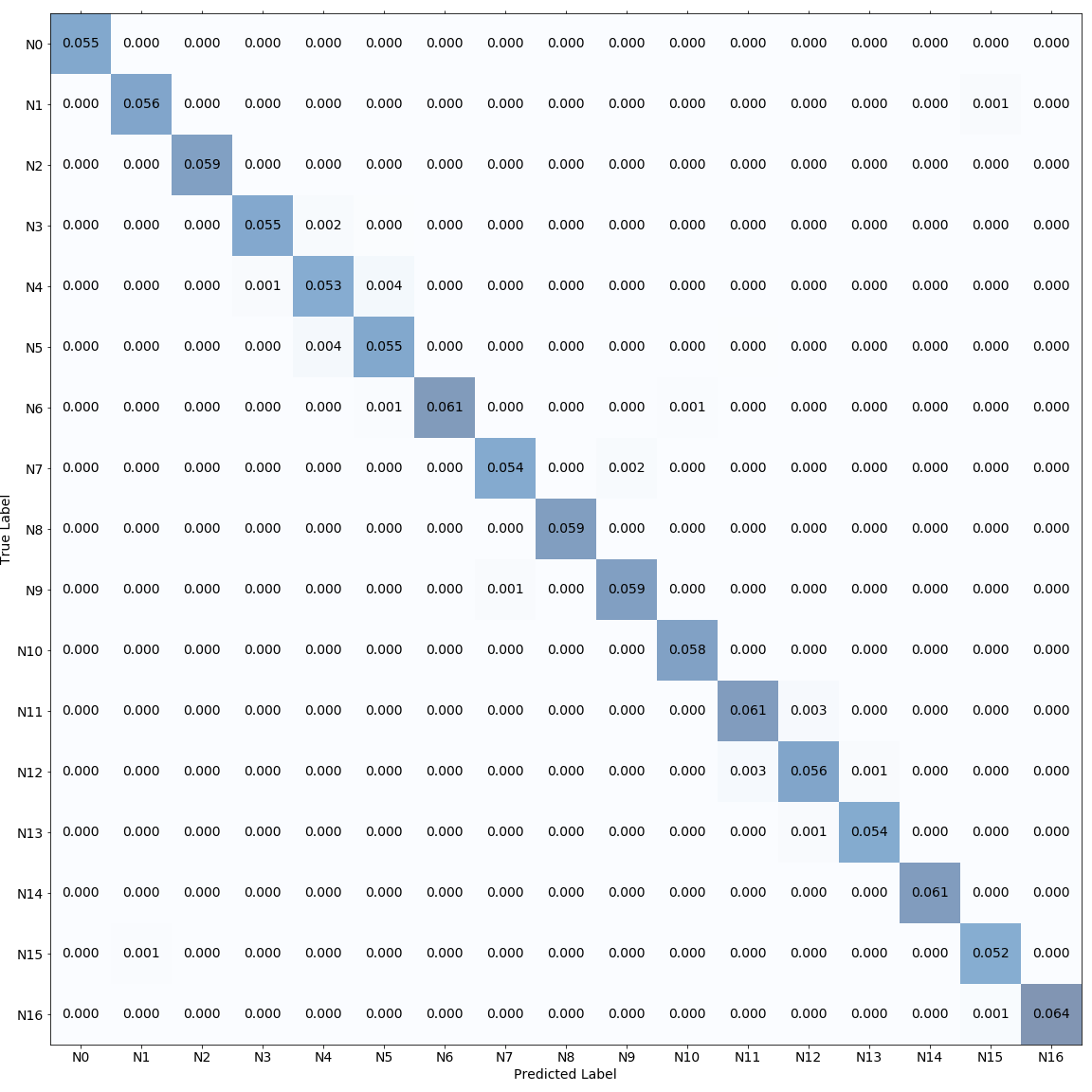}
	\caption{The confusion matrix for $17$ types of signals.}
	\label{fig:confusion17}
\end{figure}

We design a classifier to detect the difference between these signals. Using $1000$ samples for each of $17$ rotation angles, we have $17$K samples. We split the data into $80\%$ for training and $20\%$ for testing. We use patience of $8$ epochs (i.e., if loss at epoch $t$ did not improve for 8 epochs, we stop and take the best $(t-8)$ result) and train for $200$ iterations. A CNN structure similar to the one in Section~\ref{subsec:structure} is used. The only difference is that the last fully connected layer has $17$ output neurons for $17$ cases corresponding to different rotation angles (instead of $4$ output neurons). This classifier achieves $0.972$ accuracy (see Fig.~\ref{fig:17signals}-(a) for validation loss and Fig.~\ref{fig:17signals}-(b) for validation accuracy). The confusion matrix is shown in Fig.~\ref{fig:confusion17}.

\subsection{Classifier for Superimposed Signals}

\begin{figure*}
	\centering
	\subfigure[0 dB.]
	{\includegraphics[width=0.6\columnwidth]{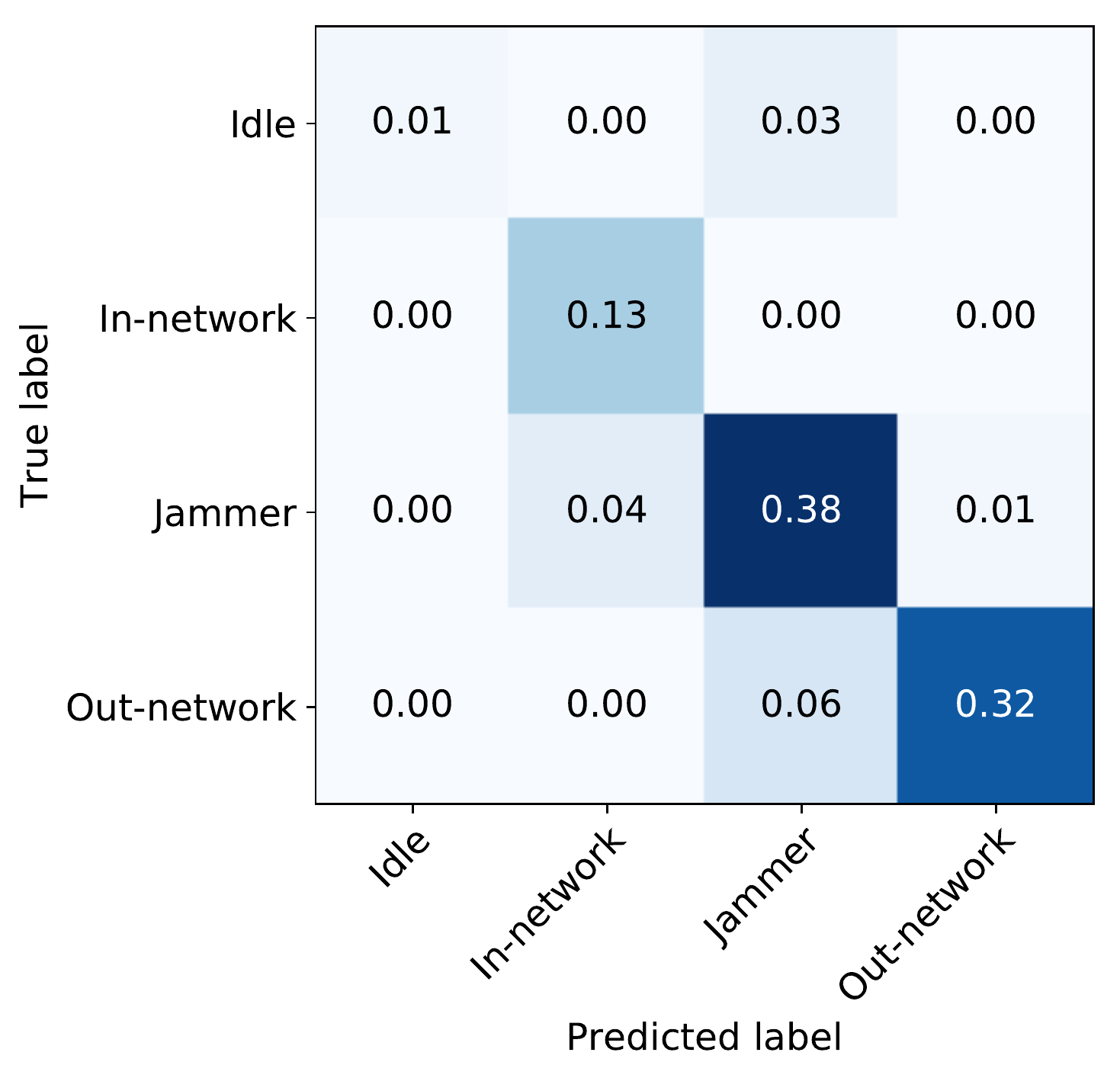}}
	\hfil
	\subfigure[10 dB.]
	{\includegraphics[width=0.6\columnwidth]{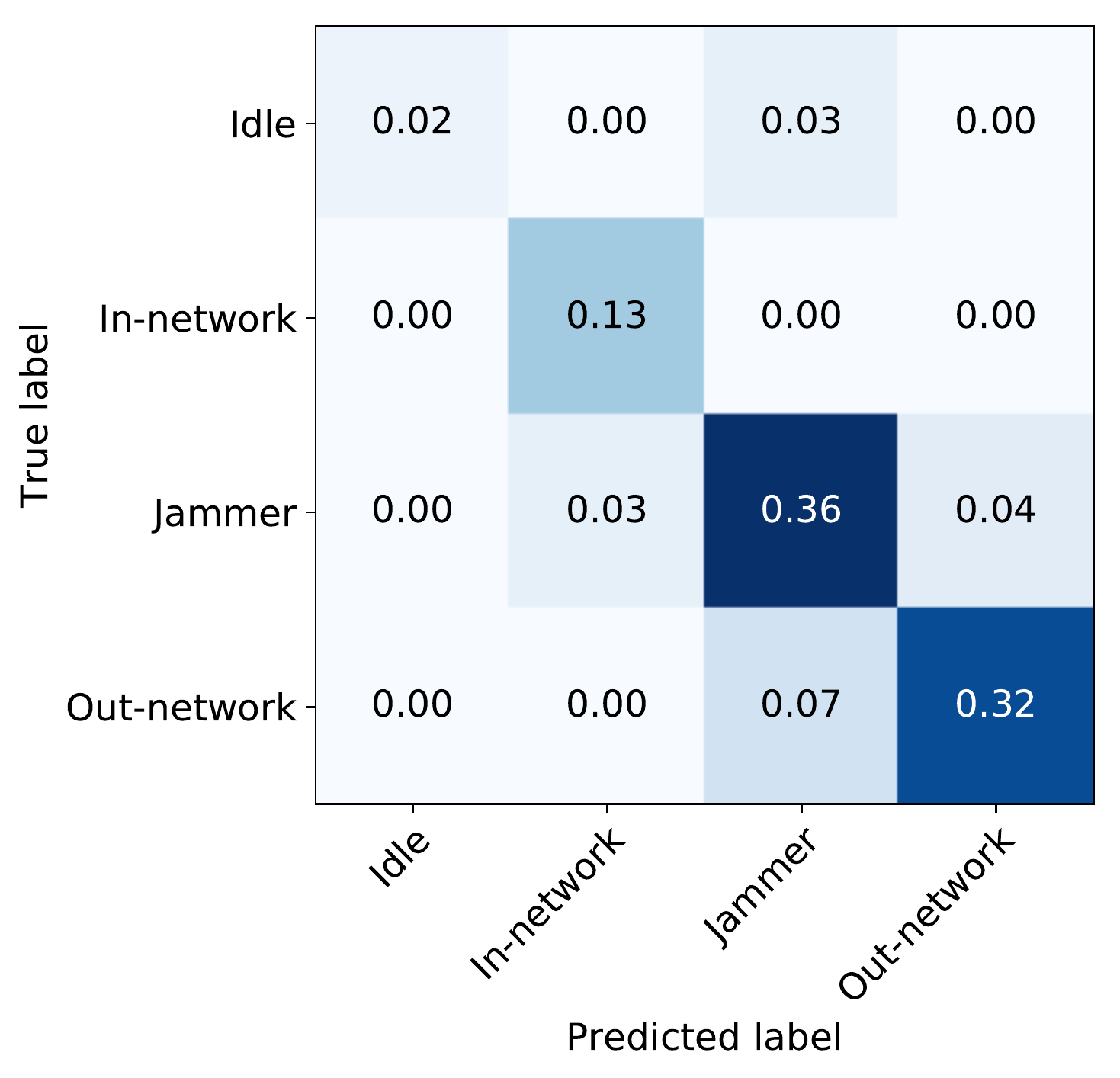}}
	\hfil
	\subfigure[18 dB.]
	{\includegraphics[width=0.6\columnwidth]{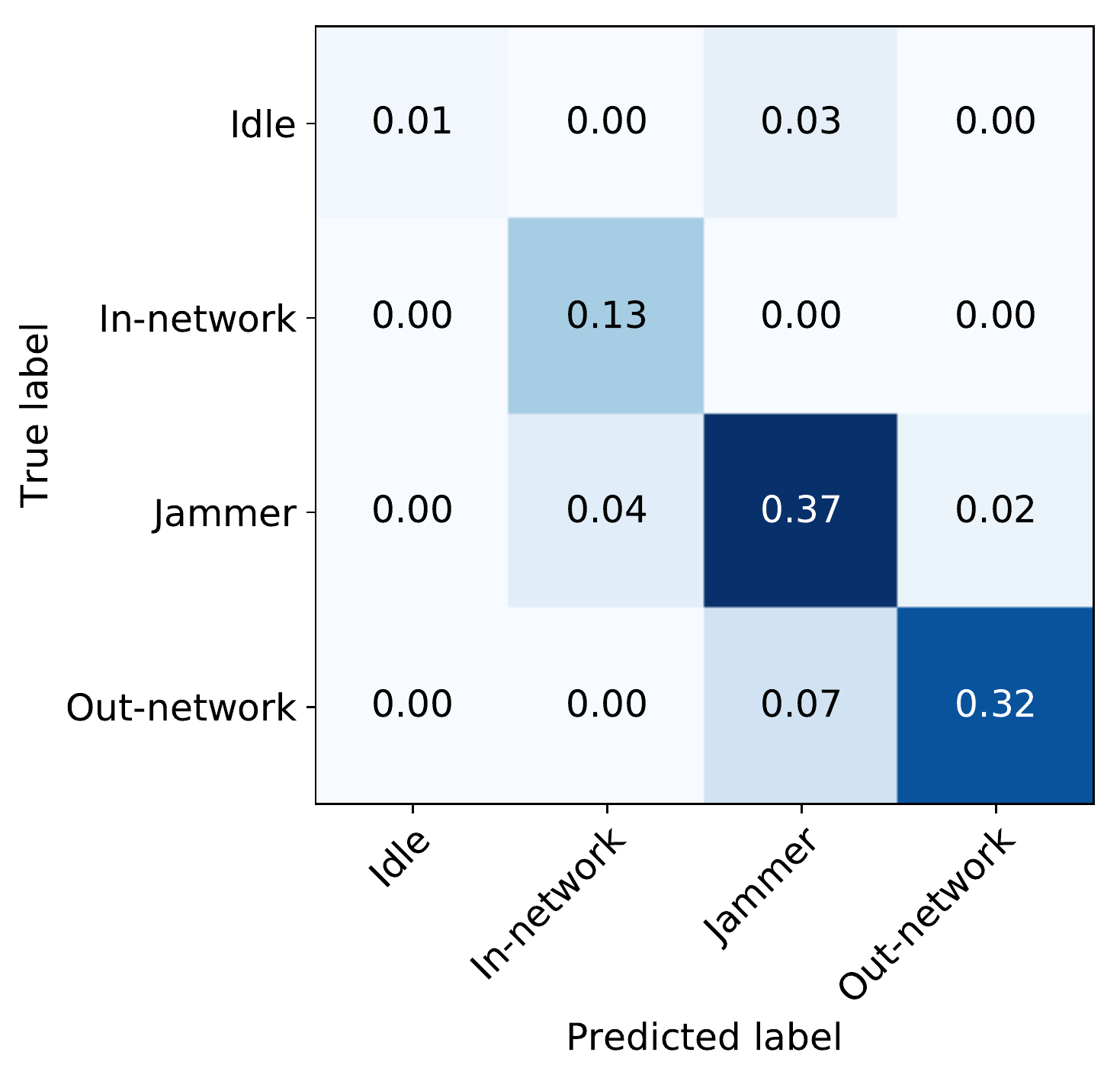}}
	\caption{Confusion matrices for superimposed signals at different SNR values.}
\label{fig:confusion4}
\end{figure*}

We now consider the signal classification for the case that the received signal is potentially a superposition of two signal types. We are particularly interested in the following two cases that we later use in the design of the DSA protocol:
\begin{itemize}
\item Superposition of in-network user and jamming signals.
\item Superposition of jamming and out-network user signals.
\end{itemize}
We first apply blind source separation using ICA. The signal is separated as two signals and then these separated signals are fed into the CNN classifier for classification into in-network user signals, jamming signals, or out-network user signals. We obtained the accuracy as shown Table~\ref{table:accuracy3} and confusion matrices at $0$dB, $10$dB and $18$dB SNR levels, as shown in Fig.~\ref{fig:confusion4}, respectively.

\begin{table}
	\caption{Accuracy for superimposed signals (averaged over all signal types).}
	\centering
	{\small
		\begin{tabular}{c|c||c|c}
			SNR (dB) & Accuracy & SNR (dB) & Accuracy \\ \hline
			0 & 0.851 & 10 & 0.824 \\ 
			2 & 0.820 & 12 & 0.834 \\
			4 & 0.857 & 14 & 0.843 \\
			6 & 0.843 & 16 & 0.830 \\
			8 & 0.827 & 18 & 0.841 \\ \hline
		\end{tabular}
	}
	\label{table:accuracy3}
\end{table}

\subsection{Classifier for Superimposed Signals}


\section{Design Distributed Scheduling Protocol}
\label{sec:scheduling}

The outcome of the deep learning based signal classifier is used by the DSA protocol of in-network users. In this section, we present a distributed scheduling protocol that makes channel access decisions to adapt to dynamics of interference sources along with channel and traffic effects.

\subsection{Superframe structure}

\begin{figure}
	\centering
	\includegraphics[width=1.0\columnwidth]{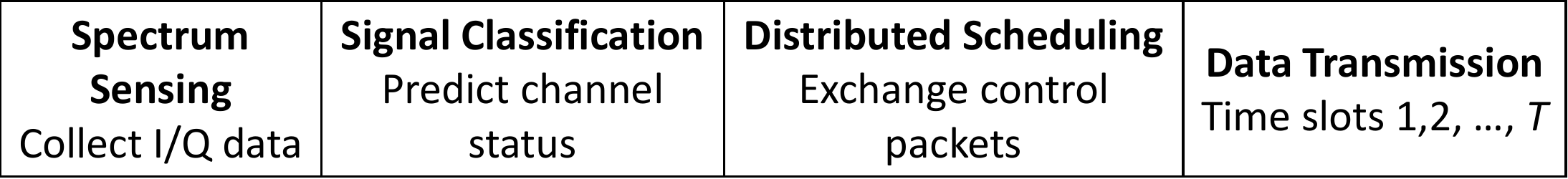}
	\caption{The superframe structure.}
	\label{fig:superframe}
\end{figure}

We consider the superframe structure (shown in Fig.~\ref{fig:superframe}) that consists of four periods:
\begin{enumerate}
\item Spectrum sensing collects I\&Q data on a channel over a sensing period.
\item Deep learning based signal classifier determines channel status based on sensing results. The status may be idle, in-network, jammer, or out-network.
\item Distributed scheduling exchanges control packages and assigns time slots to transmitters in a distributed fashion.
\item Data transmission period is divided into time slots and each transmitter sends data in its assigned time slots.
\end{enumerate}
The first three periods take a fixed and small portion of the superframe. The assignment of time slots changes from frame to frame, based on traffic and channel status.

\begin{algorithm}[t]
    \caption{Scheduling based on classification results.}
    \label{alg:scheduling}
    \begin{algorithmic}[1]
        \STATE Each receiver sends the channel status (type, score) to its transmitter if the type is not out-network. If the channel status is out-network, there should be no transmission to protect out-network user transmissions.
        \STATE If a transmitter receives channel status from its receiver and does not detect an out-network user, it broadcasts a request (type, priorities). Priorities are $T$ numbers generated based on the received score and are used to compete with other transmitters for $T$ time slots.
        \STATE A receiver generates and broadcasts a response as:
        \IF{the request type from its transmitter is smaller than the request type from all other transmitters}
        \STATE its transmitter will transmit in all $T$ time slots.
        \ENDIF
        \IF{the request type from its transmitter is the same as the request type from some other transmitters}
        \STATE its transmitter will transmit in a time slot $t$ if its priority in this time slot is the largest among these transmitters' priorities.
        \ENDIF
        \STATE A transmitter will transmit in time slot $t$ if the response from its receiver approves it to be active in time slot $t$ and responses from other receivers do not specify another transmitter in this time slot.
    \end{algorithmic}
\end{algorithm}

\subsection{Distributed Scheduling Rules}
 Scheduling decisions are made using deep learning classification results. Deep learning  provides a score on the confidence of classification to four types of signals: idle, in-network, jammer, and out-network. We use the scheduling protocol outlined in Algorithm~\ref{alg:scheduling} to schedule time for transmission of packets including sensing, control, and user data. If the in-network user classifies the received signals as out-network, it does not access the channel. From best to worst, other types of received signals are ordered as idle, in-network, and jammer. In-network users that classify received signals to better signal types gain access to channel. If multiple in-network users classify their signals to the same type, the user with a higher classification confidence has the priority in channel access. This protocol is distributed and only requires in-network users to exchange information with their neighbors.	

\subsection{Traffic Profile in Signal Classification}


Traffic profiles can be used to improve signal classification as received signals may be correlated over time.
We present next how to learn the traffic profile of out-network users and use it for signal classification. We define out-network user traffic profile (idle vs. busy) as a two-state Markov model. That is, if there is no out-network user transmission, it is in state $0$, otherwise it is in state $1$. The transition probability from state $i$ to $j$ is $p_{ij}$. Each in-network user builds its own estimation on this Markov model by online learning as follows.
\begin{enumerate}
\item Initialize the number of state changes as
\begin{eqnarray}
n_{00} =1, n_{01} =1, n_{10} =1,n_{11} =1.
\end{eqnarray}
\item Update these numbers based on past state $i$ and current predicted state $j$, i.e., $n_{ij} = n_{ij} +1$.
\item State transition probability is calculated as $p_{ij} = n_{ij}/ (n_{i0} + n_{i1})$.
\end{enumerate}

After learning the traffic profile of out-network users, signal classification results based on deep learning are updated as follows. Suppose the last status is $s_{t-1}$, where $s_{t-1}$ is either $0$ or $1$. Then based on $p_{ij}$, we can classify the current status as $s_t^T$ with confidence $c_t^T$. For example, if $s_{t-1}=0$ and $p_{00} > p_{01}$, then $s_t^T=0$ and $c_t^T=p_{00}$. Suppose the current classification by deep learning is $s_t^D$ with confidence $c_t^D$, where $s_t^D$ is either $0$ or $1$ and $c_t^D$ is in $[0.5,1]$. The classification of idle, in-network, and jammer corresponds to state $0$ in this study. We have the following three cases.
\begin{itemize}
\item $s_t^T=s_t^D$. There is no need to change classification.
\item $s_t^T=0$ and $s_t^D=1$. Then based on traffic profile, the confidence of $s_t^T=0$ is $c_t^T$ while based on deep learning, the confidence of $s_t^D=1$ is $1-c_t^D$. We use a weight parameter $w \in [0,1]$ to combine these two confidences as $wc_t^T+(1-w)(1-c_t^D)$. If this combined confidence is smaller than $0.5$, we claim that the current state is $1$, otherwise the current state is $0$. 
\item $s_t^T=1$ and $s_t^D=0$. Then based on traffic profile, the confidence of $s_t^T=0$ is $1-c_t^T$ while based on deep learning, the confidence of $s_t^D=0$ is $c_t^D$. We combine these two confidences as $w(1-c_t^T)+(1-w)c_t^D$. If this combined confidence is smaller than $0.5$, we claim that the current state is $1$, otherwise  the current state is $0$.
\end{itemize}
Note that state $0$ needs to be classified as idle, in-network, or jammer based on deep learning. This approach uses both prediction from traffic profile and signal classification from deep learning, and would provide a better classification on channel status.

\subsection{Simulation Results}
In Section~\ref{sec:sam}, the test signals are taken one by one from a given SNR. Now, we simulate a wireless network, where the SNR changes depending on channel gain, signals may be received as superposed, signal types may change over time, remain unknown, or may be spoofed by smart jammers. 

We consider the following simulation setting.
\begin{itemize}
\item $100$ in-network users are randomly distributed in a $50$m $\times 50$m region.
\item $2$ out-network users and $2$ jammers are randomly distributed in the same region.
\item Transmission/interference range is $10$m.
\item $1000$ superframes are generated. There are $10$ random links to be activated for each superframe. A superframe has $10$ time slots for data transmission.
\item Gaussian channel model is assumed.
\item If a transmission is successful, the achieved throughput in a given time slot is 1 (packet/slot).
\end{itemize}

The performance measures are in-network user throughput (packet/slot) and out-network user success ratio (\%). The goal is to improve both measures. 

\subsubsection{Benchmark Schemes} 
For comparison purposes, we consider two centralized benchmark schemes by splitting a superframe into sufficient number of time slots and assigning them to transmitters to avoid collision.
\begin{itemize}
\item Benchmark scheme 1. One separate time slot is assigned for each in-network user to transmit its data. This scheme needs $100$ time slots since there are $100$ in-network users.

\item Benchmark scheme 2. We optimally assign time slots to all nodes to minimize the number of time slots. We can build an interference graph, where each node represents a link and each edge between two nodes represents interference between two links if they are activated at the same time. If the maximum degree of this interference graph is $D$, the minimum number of time slots to avoid all interference is $D+1$.
\end{itemize}


\subsubsection{System Performance}

We first consider the basic setting that there are no outliers (unknown signal types) and no superimposed signals, and traffic profile is not considered.
The benchmark performances are given as follows.
\begin{itemize}
\item Benchmark scheme 1: In-network throughput is $760$. Out-network user success rate is $47.57\%$.
\item Benchmark scheme 2: In-network throughput is $3619$. Out-network user success rate is $47.57\%$.
\end{itemize}

\begin{table}
	\caption{System performance with different classifiers.}
	\centering
	{\small
		\begin{tabular}{c|c|c}
			Classifier & In-net user & Out-net user \\
			 & throughput & success ratio \\ \hline
			Ideal classifier (no error) & 40,578 & 100\% \\ \hline
			Random classifier & 14,563 & 80.48\% \\ \hline
			One classifier for all SNRs & 38,477 & 100\%\\ \hline
			One classifier for each SNR & 39,153 & 99.80\%
		\end{tabular}
	}
	\label{table:classifiers}
\end{table}

The performance of distributed scheduling with different classifiers is shown in Table~\ref{table:classifiers}, where random classifier randomly classifies the channel with probability $25\%$. 

We considered the effect of no jamming and obtained benchmark performance:
\begin{itemize}
\item Benchmark scheme 1: In-network throughput is $881$. Out-network user success is $47.57\%$.
\item Benchmark scheme 2: In-network throughput is $4196$. Out-network user success is $47.57\%$.
\end{itemize}

The performance of distributed scheduling with different classifiers is shown in Table~\ref{table:classifiers2}. 

\begin{table}
	\caption{System performance with different classifiers and no jamming.}
	\centering
	{\small
		\begin{tabular}{c|c|c}
			Classifier & In-net user & Out-net user \\
			 & throughput & success ratio \\ \hline
			Ideal classifier (no error) & 43,316 & 100\% \\ \hline
			Random classifier & 17,552 & 80.70\% \\ \hline
			One classifier for all SNRs & 41,780 & 100\%\\ \hline
			One classifier for each SNR & 42,614 & 99.92\%
		\end{tabular}
	}
	\label{table:classifiers2}
\end{table}

\subsubsection{Traffic profile integrated with signal classification}
We compare results with and without consideration of traffic profile, and benchmarks. We generate another instance with $p_{00} =p_{11} =0.8$ and $p_{01} =p_{10} =0.2$. The weight ($w$) to combine deep learning results and traffic profile results is set as $0.2$. We have the following benchmark performance.
\begin{itemize}
\item Benchmark scheme 1: In-network user throughput is $829$. Out-network user success is $16\%$.
\item Benchmark scheme 2: In-network user throughput is $4145$. Out-network user success is $16\%$.
\end{itemize}

The performance with and without traffic profile incorporated in signal classification is shown in Table~\ref{table:traffic}. 

\subsubsection{Classification with Outliers and Superimposed Signals}

We compare benchmark results with the consideration of outliers and signal superposition.
Benchmark performance is the same as before, since it does not depend on classification:
\begin{itemize}
\item Benchmark scheme 1: In-network user throughput is $829$. Out-network user success is $16\%$.
\item Benchmark scheme 2: In-network user throughput is $4145$. Out-network user success is $16\%$.
\end{itemize}

The performance with outliers and signal superposition included is shown in Table~\ref{table:superposition}. 

In all the cases considered, the integration of deep learning based classifier with distributed scheduling performs always much better than benchmarks.

\begin{table}
	\caption{System performance with and without using traffic profile and no jamming.}
	\centering
	{\small
		\begin{tabular}{c|c|c}
			Classifier & In-net user & Out-net user \\
			 & throughput & success ratio \\ \hline
			Without traffic profile & 40,714 & 69\% \\ \hline
			With traffic profile & 40,616 & 70\% \\
		\end{tabular}
	}
	\label{table:traffic}
\end{table}

\begin{table}
	\caption{System performance with signal superposition and no jamming.}
	\centering
	{\small
		\begin{tabular}{c|c|c}
			Classifier & In-net user & Out-net user \\
			 & throughput & success ratio \\ \hline
			Without outlier & 40,616 & 70\% \\ \hline
			With outlier & 34,236 & 69\% \\ \hline
			With superimposed signals & 34,503 & 68\% \\
		\end{tabular}
	}
	\label{table:superposition}
\end{table}

\section{Conclusion}
\label{sec:conclusion}


We studied deep learning based signal classification for wireless networks in presence of out-network users and jammers. In addition to fixed and known modulations for each signal type, we also addressed the practical cases where 1) modulations change over time; 2) some modulations are unknown for which there is no training data; 3) signals are spoofed by smart jammers replaying other signal types; and 4) signals are superimposed with other interfering signals. In case 1, we applied continual learning to mitigate catastrophic forgetting. In case 2, we applied outlier detection to the outputs of convolutional layers by using MCD and k-means clustering methods. In case 3, we identified the spoofing signals by extending the CNN structure to capture phase shift due to radio hardware effects. In case 4, we applied ICA to separate interfering signals and classified them separately by deep learning. Results demonstrate the feasibility of using deep learning to classify RF signals with high accuracy in unknown and dynamic spectrum environments. By utilizing the signal classification results, we constructed a distributed scheduling protocol, where in-network (secondary) users share the spectrum with each other while avoiding interference imposed to out-network (primary) users and received from jammers. Compared with benchmark TDMA schemes, we showed that distributed scheduling constructed upon signal classification results provides major improvements to throughput of in-network users and success ratio of out-network users.

\bibliographystyle{IEEEtran}
\bibliography{IEEEabrv,references}

\end{document}